\documentclass{iaus}
\usepackage{graphicx}
\usepackage[hyperindex,breaklinks]{hyperref}

\def\etal{{\sl et~al.\ }}
\def\spose#1{\hbox to 0pt{#1\hss}}
\def\lta{\mathrel{\spose{\lower 3pt\hbox{$\mathchar"218$}}
     \raise 2.0pt\hbox{$\mathchar"13C$}}}
\def\gta{\mathrel{\spose{\lower 3pt\hbox{$\mathchar"218$}}
     \raise 2.0pt\hbox{$\mathchar"13E$}}}

\def\=#1{\overline{#1}}

\def\cm{{\rm\,cm}}

\def\kms{{\rm\,km\,s^{-1}}}

\def\mpc{{\rm\,Mpc}}

\def\ergpscm{{\rm\,erg\,s}^{-1}{\cm}^{-2}}

\def\angs{{\rm \AA}}

\title[PNe Surveys Beyond the Local Group] 
{Planetary Nebulae Surveys Beyond the Local Group}

\author[Ortwin E.~Gerhard]   
{Ortwin Gerhard$^1$}%

\affiliation{$^1$Max-Planck-Institut f\"ur Extraterrestrische Physik,
\break Giessenbachstrasse, D-85741 Garching, Germany
\break email: gerhard@mpe.mpg.de}

\pubyear{2006}
\volume{234}  
\pagerange{1--8}
\date{3.~April 2006} 
\jname{Planetary Nebulae in our Galaxy and Beyond}
\editors{M. Barlow, R.H. Mendez, eds.}
\begin{document}

\maketitle

\begin{abstract}
Distant planetary nebulae (PNe) are used to measure distances through
the PN luminosity function, as kinematic tracers in determining the
mass distribution in elliptical galaxies, and most recently, for
measuring the kinematics of the diffuse stellar population in galaxy
clusters. This article reviews the photometric and spectroscopic
survey techniques that have been used to detect PNe beyond the Local
Group, out to the Coma cluster at 100 Mpc distance.  Contaminations by
other emission sources and ways to overcome them will be discussed as
well as some science highlights and future perspectives.
\end{abstract}

\firstsection 
\section{Introduction}

The search for planetary nebulae (PNe) beyond the Local Group is motivated
by several important science projects:

\begin{itemize}

\item The maximum luminosity of a PN in the [OIII]$\lambda
5007\angs$ line varies by at most 0.1-0.2 mag between the stellar
populations of a variety of galaxies. The PN luminosity function
(PNLF), in particular its bright cutoff, is therefore an important
distance estimator (see \cite[Jacoby 1989]{jac89}, 
\cite[Ciardullo \etal 1989]{cia89},
\cite[Jacoby, Ciardullo, Ford 1990]{jcf90},
and the review by \cite[Ciardullo 2003]{cia03}).

\item The radial velocities of PNe, also measured from the
redshifted [OIII] line, can be used to trace the kinematics of stellar
populations that have too low surface brightness for other kinematic
measurements.  PNe are therefore used as tracers of the mass
distribution in elliptical galaxies at intermediate radii ($2-5R_e$),
where measuring radial velocities from integrated absorption line
spectra is no longer feasible 
(\cite[Hui \etal 1995]{hui95},
\cite[Arnaboldi \etal 1996]{arn96}, 
\cite[1998]{arn98},
\cite[M\'endez \etal 2001]{men01}, 
\cite[Romanowsky \etal 2003]{rom03},
\cite[Peng, Ford \& Freeman 2004]{pen04}).

\item PNe are currently the only tracers (perhaps excepting
globular clusters) that allow us to measure the kinematics of the
diffuse stellar population in galaxy clusters, and thus to investigate
the dynamics and cosmological origin of this population 
(\cite[Arnaboldi \etal 2004]{arn04},
\cite[Gerhard \etal 2005]{ger05}).

\end{itemize}

\noindent
PNe have been detected out to distances of some $20\mpc$ with
narrow-band or slitless spectroscopy surveys, and out to $100\mpc$
(the Coma cluster) with multislit imaging spectroscopy (MSIS)
surveys. Sections 2-4 contain short discussions of these survey
techniques. The final Section 5 describes a few recent scientific
highlight results based on PN surveys and gives a brief outlook on
future work.

\section{Photometric surveys}\label{sec:photometric}

Optical spectra of Galactic PNe show numerous emission lines but
essentially no continuum flux. The strongest emission line is
[OIII]$\lambda5007\angs$, which contains $\sim 10\%$ of the energy
output of the central star. Further strong lines are the second
[OIII]$\lambda4959\angs$ line (three times weaker than
$\lambda5007\angs$), and H$\alpha$ ($3-5$ times weaker).  In spectra
of PNe at Virgo distance, it needs an 8m telescope to clearly see both
lines of the [OIII] doublet; such spectra only became available
recently (\cite[Arnaboldi \etal 2003]{arn03}, \cite[2004]{arn04}).

\subsection{Blinking surveys}

Based on the lack of continuum and the strong [OIII]$\lambda5007\angs$
emission line, Ciardullo, Ford, Jacoby and collaborators (e.g.,
\cite[Ciardullo \etal 1989]{cia89}, \cite[Jacoby, Ciardullo, Ford
1990]{jcf90}) developed the so-called on-band/off-band survey
technique for distant PNe.  This involves imaging the field of
interest with a narrow-band filter around the [OIII] line and with a
broad-band filter away from the dominant emission lines. In the early
surveys the PN candidates were identified by blinking the two
images. The PN candidates are those point sources in the narrow-band
image that have no counterpart in the broad-band image. Once
identified, the [OIII] fluxes of the PN candidates were determined by
aperture photometry, and converted to $m_{5007}=-2.5\log F_{5007}-13.74$
magnitudes.

As the early surveys were mostly targeted at elliptical galaxies (in
order to derive PNLF distances), contamination of the PN samples by
HII regions and background galaxies was largely absent: the surface
density of PNe in elliptical galaxies is much greater than the surface
density of background galaxies whose redshifted Ly$\alpha$ or [OII]
emission lines falls into the [OIII] filter.  To avoid contamination
by Galactic stars, the off-band image was taken to be typically $\sim
0.25$ mag deeper than the on-band image (see also \S2.2). Nonetheless,
at least one such sample contained a substantial fraction of
candidates that were not confirmed by later work (see \cite[Arnaboldi
\etal 2003]{arn03}).

The number density distributions of PN candidates in elliptical
galaxies were found to approximately follow the galaxy light,
indicating constant $\alpha_B$, the ratio of PNe to B-band luminosity.
Inside some radius, of order the effective radius $R_e$, however, the
number density profile flattens off, as fainter PNe are increasingly
lost against the bright galaxy background. This effect is present also
in more recent slitless surveys in ellipticals. Construction of the
PNLF therefore requires identification of a homogeneous and complete
PN sample, which excludes the central regions, as well as PNe fainter
than a magnitude limit set such that the remaining PNe are detected
with signal-to-noise of at least 9 per pixel (in simulations,
essentially all objects are detected at this S/N level).

\subsection{Wide-field surveys}

Following the discovery of three intergalactic PNe in the
Virgo cluster near M86 (\cite[Arnaboldi \etal 1996]{arn96}), wide-field
imaging (WFI) surveys were undertaken to map this much more sparse PN
population in Fornax, Virgo, and several galaxy groups
(\cite[Feldmeier,  Ciardullo \& Jacoby 1998]{fel98},
\cite[Arnaboldi \etal 2002]{arn02},
\cite[2003]{arn03},
\cite[Ciardullo \etal 2002]{cia02a},
\cite[Feldmeier \etal 2003]{fel03},
\cite[Castro-Rodr{\'{\i}}guez \etal 2003]{cas03},
\cite[Aguerri \etal 2005]{agu05}).
Again, images are taken in a narrow-band filter around the red-shifted
[OIII]$\lambda5007\angs$ line, typically $\simeq 60\angs$ wide to
cover the wider cluster velocity distribution, and in a nearby broad-band
filter. 
Such WFI fields contain many 1000's of stars, so the blinking method
is no longer feasible and automatic methods for sifting out the PNe
must be used. One such technique is based on the narrow-band vs.\
broad-band colour magnitude diagram (CMD), as pioneered by 
\cite{thw97}. An example for the Virgo RCN1 field from \cite{arn02}
is shown in Figure~1. PNe candidates in this field are those to the
right of the diagonal line indicating the continuum limit, and 
below the other lines to exclude Galactic stars and background 
emission galaxies with EW below 110\angs, as described in the figure
caption.
Because of the much lower surface density of PNe in these intracluster
fields, the PN catalogues extracted from the images may contain a
substantial fraction of contaminants. We now briefly discuss these
various interlopers and how to avoid them.

\begin{figure}
\centering
\includegraphics[width=.8\textwidth]{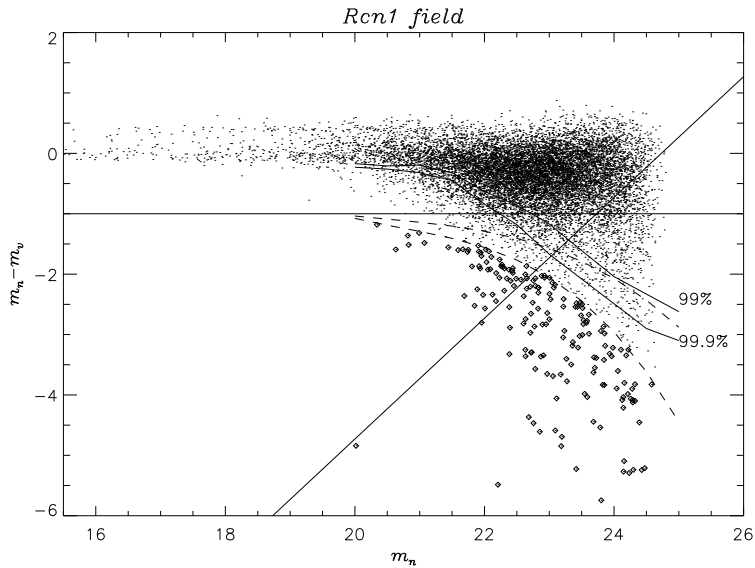}
\caption{Color-magnitude diagram for all the unresolved sources in 
the Virgo RCN1 field. The horizontal line at $m_{n}\!-\!m_{v}=-1$
indicates objects with an observed EW=110 \angs. The diagonal line shows
the magnitude corresponding to $1.0 \times \sigma$ above sky in the V
band. Full curved lines represent the 99$\%$ and 99.9$\%$ lines for
the distribution of modelled continuum objects. The dashed lines
represent 84\% and 97.5\% lines for the distribution of modelled
objects with $m_n\!-\!m_v=-1$. The points are all objects detected by
SExtractor. Diamonds are objects with a significant color excess in
the narrow band filter. From \cite{arn02}.
\label{fig1}}
\end{figure}

{\bf Ordinary Galactic stars.} In Fig.~1, stars were excluded from the
PN catalogue by considering their $m_n-m_b$ colour distribution after
convolution with the photometric errors and discarding all objects
above the 99.9\% line. However, the photometric errors may conspire to
result in a bright narrow-band $m_n$ and faint broad-band $m_b$, so as
to result in apparently large EW (negative colour) and undetectable
continuum, even for stars with AB colour zero (requiring sufficiently
large negative error in $m_n$ and positive error in $m_b$). Such
errors move stars from the top right of Fig.~1 into the region of the
diagram containing the high-EW sources. While statistically not very
likely, for $\sim 10^4$ stars in the image this ``spill-over'' effect
(\cite[Aguerri \etal 2005]{agu05}) may occur often enough to lead to a
significant contamination of the PN samples.  From the simulations for
their fields \cite{agu05} found that 60\%/20\%/10\% of their Virgo
intracluster PN (ICPN) samples were stars due to the spillover effect,
if the limiting magnitude of the off-band image was 0/0.5/0.9 mag
deeper than the limiting magnitude of the on-band image. Thus the
contamination by faint stars may be avoided by using sufficiently deep
off-band images.

{\bf HII regions.} Imaging surveys including H$\alpha$ have found a
few compact HII regions in one Virgo field (\cite[Gerhard \etal
2002]{ger02}, \cite[Arnaboldi \etal 2003]{arn03}), probably
associated with gas that was stripped recently from galaxies, but this
is a small fraction of all sources in the ICPN catalogue.

{\bf [OII] emission galaxies} at redshift $\simeq0.35$, whose [OII]
emission lines fall into the [OIII] narrow-band filter at Virgo
redshift.  The contamination from these objects can be greatly reduced
by retaining only emission sources with observed EW greater than
$100\angs$: \cite{ham97} and \cite{hog98} found no [OII] emitters with
restframe EW $> 70\angs$. This has been confirmed by spectroscopic
follow-up observations of Virgo ICPNe candidates, none of which showed
the [OII] doublet and a continuum (\cite[Arnaboldi \etal
2004]{arn04}).  Of course, such a stringent EW cut also eliminates
some PNe from the catalogue, for which the emission is too weak
compared to the sky noise.

{\bf Lyman $\alpha$ emitters} at redshift $\simeq3.1$, whose
Ly$\alpha$ line also comes into the [OIII] filter. This class of
contaminants was found originally by \cite{kud00} taking spectra of
ICPN candidates in a Virgo field which indeed is consistent with
containing no intracluster light (ICL; \cite[Aguerri \etal
2005]{agu05}, \cite[Mihos \etal 2005]{mih05}).  Subsequent studies
of Ly$\alpha$ emitters in blank fields have shown that the Ly$\alpha$
luminosity function is spatially variable and that the brightest
Ly$\alpha$ emitters are between $0.5-1.5$ magnitudes fainter than the
PNLF cutoff in Virgo (\cite[Ciardullo \etal 2002b]{cia02b},
\cite[Castro-Rodr{\'{\i}}guez \etal 2003]{cas03}, \cite[Aguerri \etal
2005]{agu05}). With good spatial resolution extended Ly$\alpha$
galaxies may be eliminated, but according to the simulations of
\cite{agu05} typically $\sim 25\%$ of the ICPN candidates in Virgo
could be Ly$\alpha$ background galaxies.

A large fraction of these will be recognized in follow-up spectroscopy
by their continuum, their broad emission lines, or by the absence of
[OIII]$\lambda4959\angs$ if this is in the spectral range. The
spectroscopic results of \cite{arn04} have confirmed $80\%$ of the
expected number of PNe in the recent Virgo photometric surveys as real
PNe, showing that these surveys are now well-understood.  Many of the
confirmed Virgo ICPNe also show the second [OIII]$\lambda4959\angs$
line besides [OIII]$\lambda5007\angs$, confirming without doubt that
these objects are in the cluster.

The measured radial velocities then provide us with the kinematics of
the intracluster stellar population; see the scientific highlight
Section \ref{sec:highlights}. Similarly, the combination of
photometric survey plus spectroscopic follow-up has been used to
obtain samples of PN radial velocities in the halos of elliptical
galaxies. See the references in the Introduction.

\section{Slitless spectroscopy surveys}\label{sec:slitless}

\subsection{Dispersed/undispersed imaging}

In this technique pioneered by \cite{men01}, undispersed images of the
field are taken both through a broad band filter and through a narrow
band filter centered around the redshifted [OIII]$\lambda5007\angs$
line for the target PNe. The PN candidates are identified as those
sources that appear only in the on-band image, which also provides
accurate astrometric positions.  Additionally, a dispersed image is
taken through a grism and the on-band filter.  In this image, the PN
candidates are still point sources, but are now shifted relative to
their true positions on the sky by an amount reflecting both the
mapping by the grism and their radial velocities.  By contrast, stars
in the field now appear as elongated features on the CCD, with length
corresponding to the filter width.  

From such data in two overlapping fields, \cite{men01} obtained fluxes
and radial velocities for 535 PN candidates in the E5 galaxy NGC 4697,
with velocity errors of $\sim 40\kms$ and magnitude errors of
$0.1-0.2$ mag for the bright/faint sources.  Because this is a
spectroscopic technique, there are fewer contaminants than in the
imaging surveys; only background galaxies with faint continuum and
narrow emission line will remain unrecognized. 
\cite[M\'endez \etal]{men01} obtained a new PNLF distance to NGC 4697
of $D\simeq 10.5\mpc$, and found that the falling velocity dispersion
profile in this galaxy is consistent with simple isotropic spherical
models without dark matter halo out to $\sim 2.5R_e$.

A similar analysis was done by \cite{teo05} for the Fornax elliptical
galaxy NGC 1344. However, in this galaxy the velocity dispersion
profile is not consistent with isotropic constant-M/L models, so most
likely some dark matter is indicated. For more details on this method
and the results obtained, see M\'endez, this conference.

\subsection{Counter-dispersed imaging (CDI)}

The CDI technique instead uses two images dipersed in opposite
directions, to determine both the true positions of the PN candidates
on the sky and their radial velocities. This technique is employed by
the special-purpose PN.S instrument mounted at the WHT at La Palma, in
which the incoming beam, after passing the [OIII]$\lambda5007\angs$
filter, is split into two parts that are simultaneously imaged through
separate grisms in two [OIII] arms. (A third H$\alpha$ arm is
currently built that will also allow simultaneous H$\alpha$ imaging of
the field).  Two noteworthy properties of the PN.S instrument are its
wide ($11.'4\times 10.'3$) field of view and high system efficiency
($33\%$); see \cite[Douglas \etal (2002)]{dou02}. With the H$\alpha$
arm, it will also be possible to remove high-redshift galaxies with
narrow emission line and faint continuum, which are the remaining
contaminants in the current samples, similar as in the \cite{men01}
technique.

First results based on PN.S data for three intermediate-luminosity
elliptical galaxies were published by \cite[Romanowsky \etal
(2003)]{rom03}.  Like \cite{men01} for NGC 4697, they found that all
three galaxies (NGC 821, 3379, 4494) have falling dispersion profiles
out to several $R_e$, which are consistent with dynamical models with
little if any dark matter out to these radii. The PN.S has also been
highly successful in producing a sample of some 2500 PNe in M31
(\cite[Merrett \etal 2006]{mer06}).  Currently, most of the data for
the PN.S team's core science program have been taken, and are being
analyzed with an improved data pipeline and revised dynamical
models. The goal of the core science program is to study a sample of
12 round elliptical galaxies and obtain constraints on their dark
matter halos out to $\sim 6 R_e$ (for more details, see Romanowsky,
this conference).

\section{Multi-slit imaging spectroscopy surveys}\label{sec:msis}

Beyond about $20-30\mpc$ distance, PNe are too faint to be detectable
with narrow band surveys or slitless spectroscopy - their emission
disappears in the sky noise in the narrow band filter. The brightest
PNe in the Coma cluster at $100\mpc$ distance have line fluxes of
$2.2\times 10^{-18} \ergpscm$ -- this is equivalent to $\sim 20$
photons per minute through the aperture of an 8m telescope, of which
$\sim 2$ will reach the detector for a typical $\sim 10\%$ overall
system efficiency.  To detect PNe at such distances requires a
spectroscopic blind search technique: spectroscopic, so that only the
sky noise within a few \angs\ dilutes the emission from the PN, and
blind, because the positions of these faint PNe cannot be previously
determined.

The Multi-Slit Imaging Spectroscopy (MSIS) technique (\cite[Gerhard
\etal 2005]{ger05}) which meets these requirements combines a mask of
parallel multiple slits with a narrow-band filter, centered on the
redshifted [OIII]$\lambda5007\angs$ emission line.  Spectra are
obtained of all PNe that lie behind the slits. The narrow band filter
limits the length of the spectra on the CCD so that many slits can be
simultaneously exposed.  For each set of mask exposures only a
fraction of the field is surveyed; to increase the sky coverage the
mask can be stepped on the sky. The approach is similar as in some
searches for high $z>5$ Ly$\alpha$ emitters (e.g., \cite[Tran \etal
2004]{tran+04}).

In their pilot study, \cite{ger05} obtained MSIS observations of the
\cite{Bern+95} field in the Coma cluster core. This field is at the
center of the cluster X-ray emission and $\sim5$ arcmin away from the
cD galaxy NGC 4874. From the \cite{Bern+95} diffuse light
measurement we expected about 425 PNe in this field. The fraction of
the field surveyed by the mask was $\sim12\%$; because $\sim$ half of
the PNe will be dimmed as they are not centered in the slits, we
estimated to detect $\sim20\!-\!30$ PNe per mask.

\begin{figure}
\setlength{\unitlength}{0.01\textwidth}
\begin{picture}(50,62)
 \put(0,2)  {\includegraphics[width=48\unitlength]{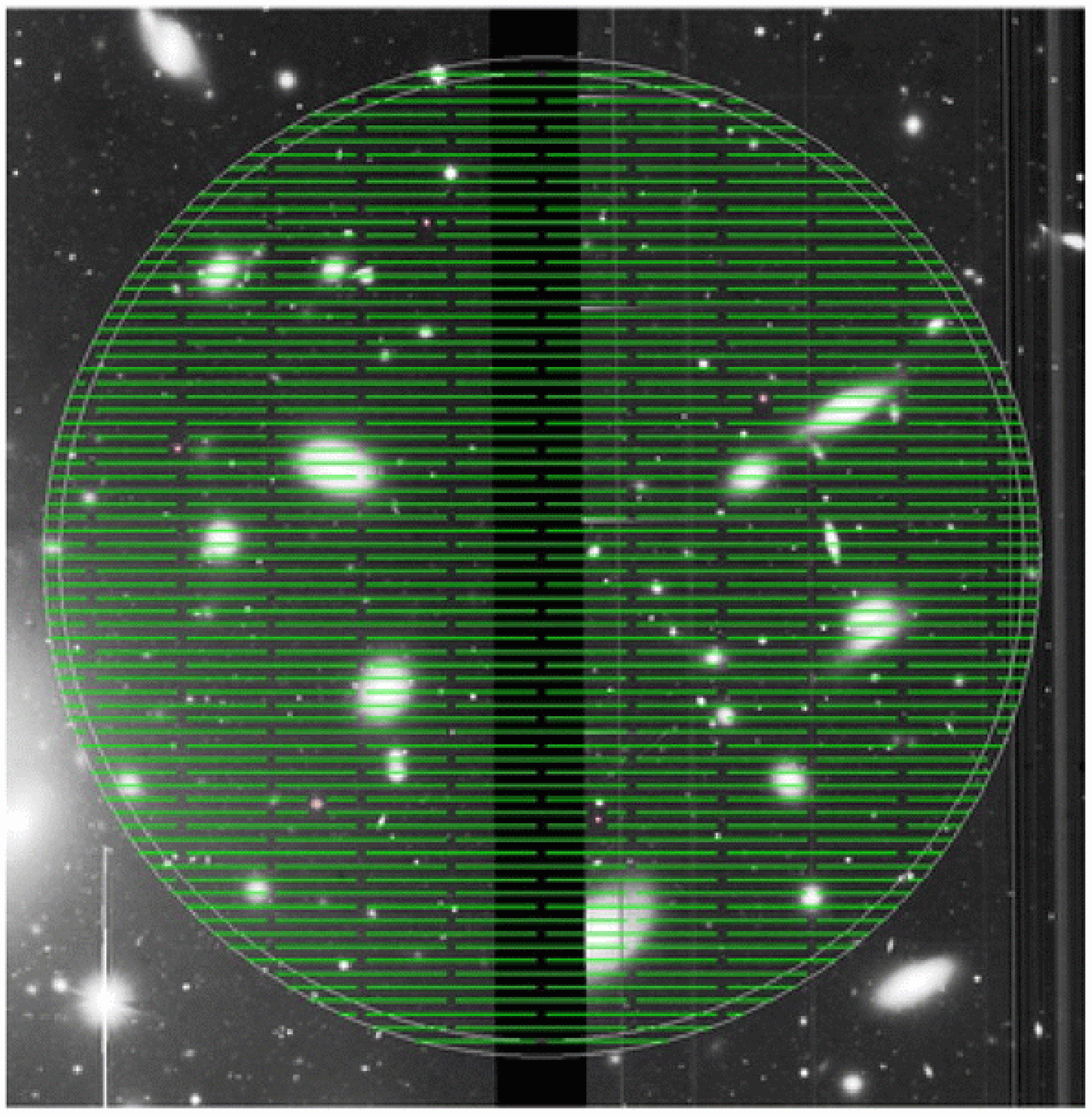}}
 \put(52,40) {\includegraphics[width=48\unitlength,height=22\unitlength]{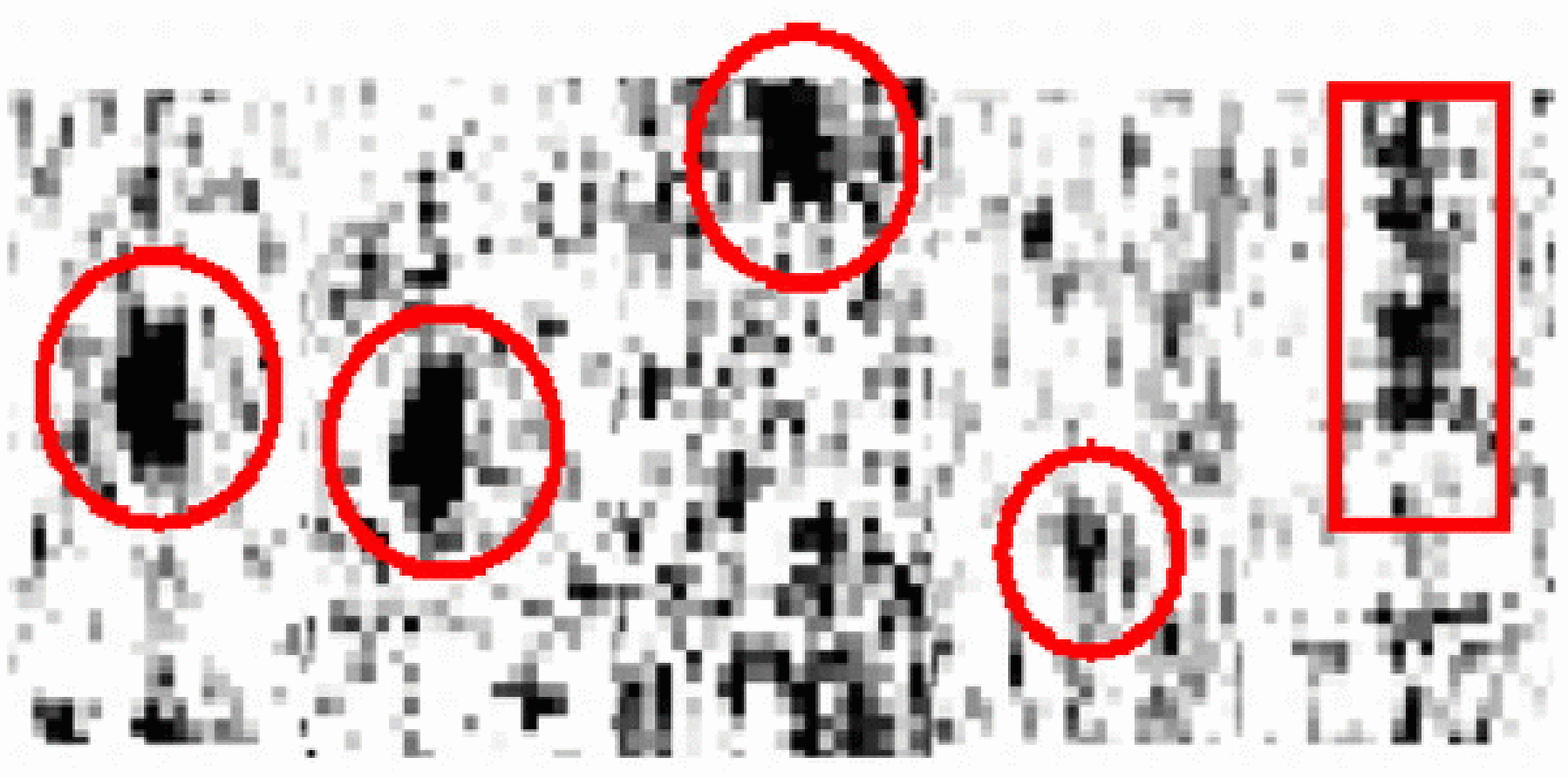}}
 \put(49,0){\includegraphics[width=51\unitlength,height=40\unitlength]{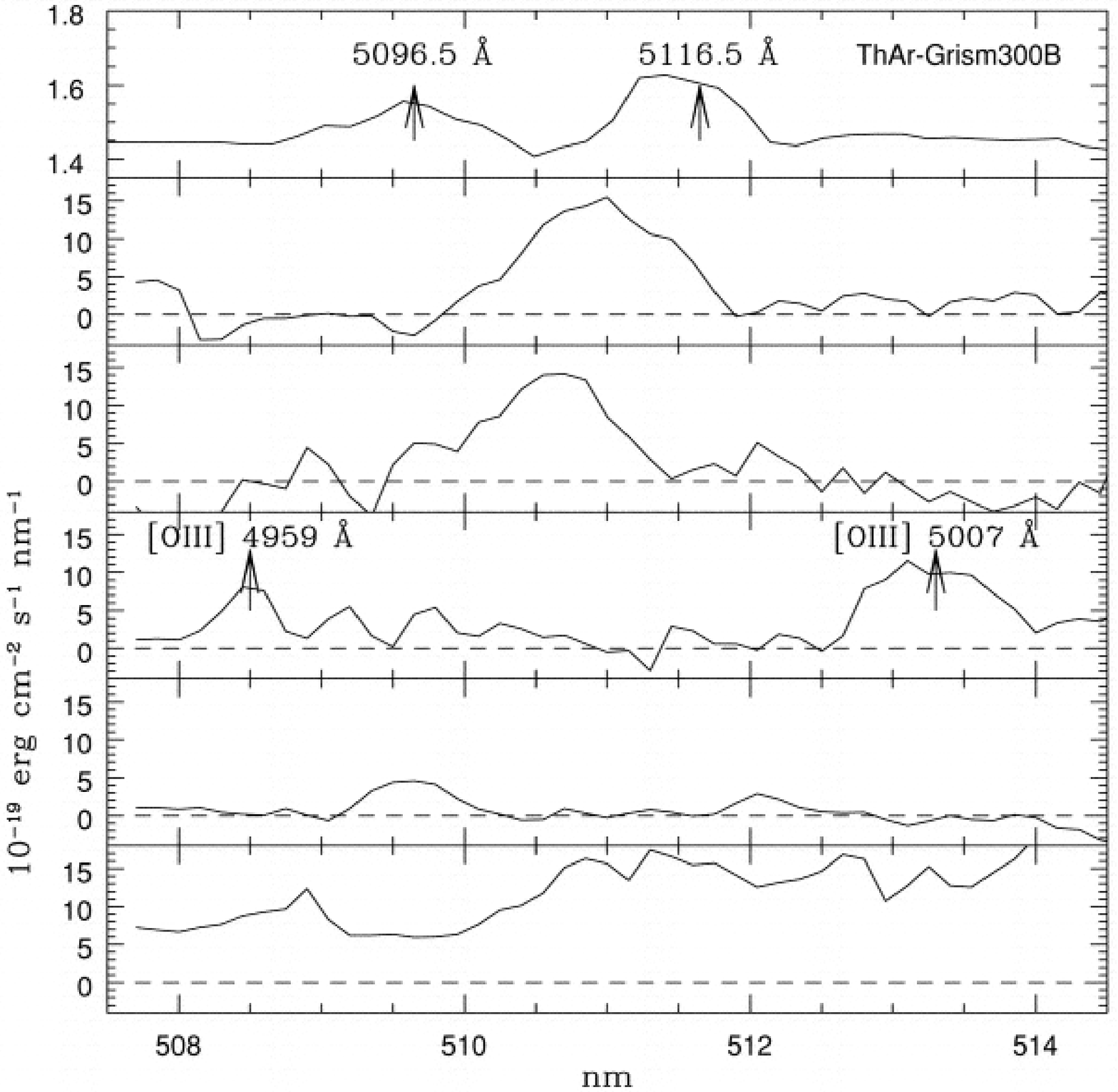}}
\end{picture}
\caption{Left: Multi-slit mask used in the Coma MSIS observations,
superposed on the Coma core field.
Right: Two-dimensional (top) and one-dimensional (bottom) spectra
of 4 PN candidates and a probable Ly$\alpha$ galaxy (right and bottom
panel, respectively) in the Coma core field. 
Also shown is a 1D arc lamp spectrum at the spectral resolution of the setup,
$7.2\angs=440\kms$. Note the second [OIII]$\lambda4959\angs$ line in the third
PN spectrum. After \protect\cite{ger05}.
[See online paper for colour version.]
\label{fig2}}
\end{figure}

In the MSIS setup, a compromise must be found between the number of
slits and hence number of objects detected, and the velocity
resolution and signal-to-noise (S/N) of the emission sources, which
are dependent on the spectral resolution. For the Coma observations, a
mask with 70 slits was used, each spectrum was 43 pixels on the CCD,
and the spectral resolution $d$ corresponded to $440\kms$.  Then the
total S/N over all pixels becomes
\newcommand{\SNRPN}{ {\rm
SNR}_{PN} }
\begin{equation}
 \SNRPN = 7.0 \times 10^{-0.4\Delta m}
 \times
\left(\frac{d}{1.45{\mbox \angs}/{\rm pix}}\right)^{-1/2}
\left(\frac{\phi}{0.''6}\right)^{-1}
\left(\frac{\theta}{6\sigma_y}\right)^{-1/2}
\left(\frac{t_{exp}}{3~hrs}\right)^{1/2}
\end{equation}
for a PN fainter than the cutoff by $\Delta m$ magnitudes.  Here
$\phi$ is the seeing FWHM, $\theta$ the slit width, $\sigma_y$ the
pixel size, and $t_{exp}$ the exposure time.  Relative to the sky
noise in one pixel, the corresponding S/N of a PN at the PNLF cutoff
$(\Delta m=0)$ was 32. Both values assume that the emission from the
PN falls through the slit completely.  Detection of a source is
normally considered secure if the S/N relative to the sky noise in one
pixel is $\ge 9$. In the Coma data this corresponds to $\SNRPN \ge 2$,
i.e., $95\%$ probability of the source being real. If this criterion
is used for secure detection, PNe about 1.4 magnitudes down the PNLF
could be detected. It is clear from equation~(4.1) that the most
important condition for successful detection is good seeing.

Figure~2 shows the mask used in the Coma observations, as well as the
two-dimensional and one-dimensional spectra of some of the PN
candidates and a probable Ly$\alpha$ emitting galaxy, and of the arc
lamp. The faintest candidate in Fig.~2 has a total $\SNRPN = 2.0$,
corresponding to a flux of $3\times10^{-19}\ergpscm$ ($\simeq 68$
photons in three hours exposures with FOCAS). The reasons to believe
that these emission sources are indeed almost all PNe in the Coma
cluster are: (1) They are unresolved spatially and in wavelength; (2)
they have no detectable continuum, down to
$1.6\times10^{-20}\ergpscm\angs^{-1}$, the flux from two O stars at
the distance of Coma; (3) for all four bright sources for which
[OIII]$\lambda5007\angs$ is sufficiently redshifted, the second
[OIII]$\lambda4959\angs$ line has been detected in the blue part of
the filter; (4) the emission fluxes are consistent with those expected
for PNe in Coma, at and below the bright cutoff of the PNLF; (5) the
number density and (6) the radial velocity distribution are also
consistent with the surface brightness of ICL and the radial
velocities of Coma galaxies in this field. On the contrary, emission
sources with measured continuum (likely background galaxies) have a
much more uniform velocity distribution.

These first ICPN observations in Coma have already lead the way to the
conclusion that the two cD galaxies with their associated subclusters
are currently undergoing a strong interaction prior to merging; see
the talk by Arnaboldi in this conference. A lot more insight on the
dynamics of present-day clusters and their ICL can be expected in the
next few years -- with the MSIS technique, velocities of ICPNe can 
be measured in all clusters out to $\sim 100\mpc$ distance.

\section{Some recent science highlights and future 
prospects}\label{sec:highlights}

Here I will briefly mention three areas in which recent PN survey work
has led to important and sometimes surprising results, and where
further work will be needed.

{\bf (1) PNe population effects.} A reanalysis of the \cite{men01}
sample of 535 PNe in the elliptical galaxy NGC 4697 has shown that
most of the brightest of these PNe are part of a subpopulation which
is azimuthally unmixed and kinematically peculiar, and thus neither
traces the distribution of all stars nor can be in dynamical
equilibrium in the gravitational potential (\cite[Sambhus \etal
2006]{sam06}).  This introduces an uncertainty in the PNLF cutoff
luminosity for this galaxy of $\sim 0.15 {\rm mag}$; the PNLF of two
kinematic subsamples differ with statistical significance. Probably
related is the recent result that the bright and faint ICPNe in the
MSIS Coma cluster sample (see \S\ref{sec:msis}) have different
velocity distributions (Arnaboldi \etal, in preparation). 

{\bf (2) Low density dark matter halos in ellipitical galaxies.} The
work of \cite{men01} and \cite{rom03} based on slitless spectroscopy
PN surveys has provided evidence that in some intermediate luminosity
ellipticals the velocity dispersion profiles are falling out to at
least several $R_e$, suggesting that the dark matter halos in these
galaxies must be fairly diffuse. However, other ellipticals appear to
have denser halos (\cite[e.g., Gerhard \etal 2001]{ger01}); also,
further study is needed to address the concerns of \cite{dek05}. The
PN.S key project results will be very valuable to clarify these
issues.

\begin{figure}
\includegraphics[bb=20 120 570 790,clip=true,angle=-90,width=\textwidth]{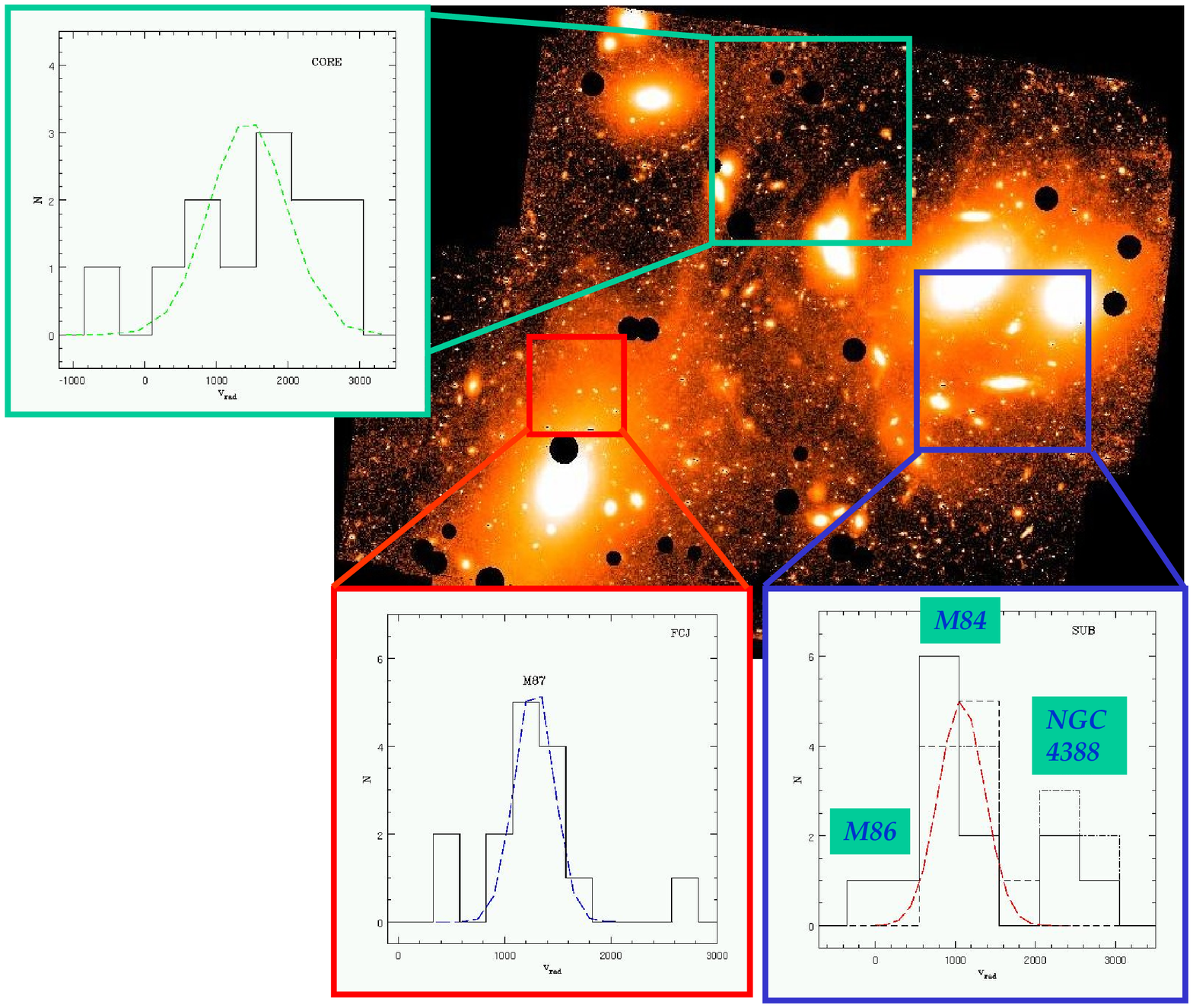}
\caption{Montage of a deep image of the Virgo cluster core from
\cite{mih05}, including M87 (lower left) and M86/M84 regions (right),
and histograms of ICPN velocities measured by \cite{arn04} in the
indicated fields. The image and kinematics show that M87 has an
extended, dynamically bound halo. By contrast, the large halo around
M84, M86, and NGC 4388 is a superposition; the PNe velocities in this
field can clearly be associated with the individual galaxies, but M84
also has an extended, dynamically bound halo. The streamers in the
image as well as the ICPN velocities in the Core and M87 fields show
that the intracluster stars in Virgo have not yet mixed to virial
equilibrium.
[See online paper for colour version.]
}
\end{figure}

{\bf (3) Kinematics of ICL.} Figure 3 based on the results of
\cite{arn04} and \cite{mih05} gives a vivid illustration how the
velocities of ICPNe can shed light on the kinematics and dynamics of
the ICL, showing clearly the dynamical youth of the Virgo
cluster. Both M87 and M84 have extended stellar halos in approximate
dynamical equilibrium with their parent galaxies. Superposed on the
M87 halo and in the CORE field stars are seen which clearly have
intracluster kinematics but whose velocity distributions do not appear
to be well-mixed yet, and in the SUB field, all of the PNe observed
are consistent with belonging dynamically to the large galaxies in the
field. ICPN and galaxy velocities indicate that, somewhat
surprisingly, also the much denser Coma cluster is in a stage of
strong dynamical evolution (\cite[Gerhard \etal 2005]{ger05},
Arnaboldi \etal, in preparation, \cite[Adami \etal
2005]{ada05}). Further measurements of ICPN velocities as discussed in
\S4 will show whether generally the ICL in nearby galaxy clusters is
inhomogeneous and dynamically evolving, or whether there are clusters
in which it has already found a reasonably mixed equilibrium
configuration.
  
\begin{acknowledgments}
I would like to acknowledge the Swiss National Science Foundation
for supporting the research program on PNe in distant galaxies and
clusters during several grant periods.
\end{acknowledgments}


\newpage

\begin{thebibliography}{}

\bibitem[Adami \etal (2005)]{ada05} Adami, C., Biviano, A., Durret,
F., Mazure, A., 2005, \textit{AA}, 443, 17

\bibitem[Aguerri \etal (2005)]{agu05} Aguerri, J.A.L., Gerhard, O.E., 
Arnaboldi, M., Napolitano, N., \etal 2005, \textit{AJ}, 129, 2585

\bibitem[Arnaboldi  \etal (2002)]{arn02} Arnaboldi, M., Aguerri, J.A.L.,
Napolitano, N.R., Gerhard, O., \etal, 2002, \textit{AJ}, 123, 760

\bibitem[Arnaboldi \etal (1998)]{arn98} Arnaboldi, M., Freeman, K.C., 
Gerhard, O., Matthias, M., \etal 1998, \textit{ApJ}, 507, 759

\bibitem[Arnaboldi \etal (1996)]{arn96} Arnaboldi, M., Freeman, K.C.,
Mendez, R.H., Capaccioli, M., \etal 1996, \textit{ApJ}, 472, 145

\bibitem[Arnaboldi \etal (2003)]{arn03} Arnaboldi, M., Freeman, K.C.,
Okamura, S., Yasuda, N., \etal, 2003, \textit{AJ}, 125, 514  

\bibitem[Arnaboldi \etal (2004)]{arn04} Arnaboldi, M., Gerhard, O.E.,
Aguerri, J.A.L., \etal, 2004, \textit{ApJ} (Letters), 614, L33

\bibitem[Bernstein \etal (1995)]{Bern+95}
Bernstein, G.M., Nichol, R.C., Tyson, J.A., Ulmer, M.P., Wittman, D.,
1995, \textit{AJ}, 110, 1507

\bibitem[Castro-Rodr{\'{\i}}guez \etal (2003)]{cas03} 
Castro-Rodr{\'{\i}}guez, N., Aguerri, J.A.L., Arnaboldi, M., 
Gerhard, O., \etal, 2003, \textit{AA}, 405, 803 

\bibitem[Ciardullo(2003)]{cia03} Ciardullo, R. 2003, in Stellar Candles
for the Extragalactic Distance Scale, ed. D. Alloin \& W. Gieren, Lect
Notes Phys., 635, 243

\bibitem[Ciardullo \etal (2002a)]{cia02a} Ciardullo, R.,  
Feldmeier, J.J., Krelove, K., Jacoby, G.H., Gronwall, C., 2002a, 
\textit{ApJ}, 566, 784  

\bibitem[Ciardullo et al. (2002b)]{cia02b} Ciardullo, R., Feldmeier, J.J.,
Jacoby, G.H., Kuzio de Naray, R., \etal, 2002b, \textit{ApJ}, 577, 31

\bibitem[Ciardullo \etal (1989)]{cia89} Ciardullo, R., Jacoby, G.H., Ford, 
H.C. 1989, \textit{ApJ}, 344, 715 

\bibitem[Dekel \etal (2005)]{dek05} Dekel, A., Stoehr, F., Mamon, G.A., 
Cox, T.J., Novak, G.S., \etal, 2005, \textit{Nature}, 437, 707

\bibitem[Douglas \etal (2002)]{dou02} Douglas, N.G., Arnaboldi, M.,
Freeman, K.C., Kuijken, K., \etal, 2002, \textit{PASP}, 114, 1234

\bibitem[Feldmeier,  Ciardullo, \&  Jacoby (1998)]{fel98}  Feldmeier, J.J., 
Ciardullo, R., Jacoby, G.H., 1998, \textit{ApJ}, 503, 109 
 
\bibitem[Feldmeier \etal (2003)]{fel03} Feldmeier, J.J., Ciardullo, R., 
Jacoby, G.H., Durrell, P.R., 2003, \textit{ApJS}, 145, 65
 
\bibitem[Gerhard \etal (2002)]{ger02} Gerhard, O., Arnaboldi, M.,
Freeman, K.C., Okamura, S., 2002, \textit{ApJ} (Letters), {580}, {L121}

\bibitem[Gerhard \etal (2005)]{ger05} Gerhard, O., Arnaboldi, M.,
Freeman, K.C., \etal, 2005, \textit{ApJ} (Letters), {621}, {L93}

\bibitem[Gerhard \etal (2001)]{ger01} Gerhard, O., Kronawitter, A.,
Saglia, R.P., Bender, R., 2001, \textit{AJ}, {121}, {1936}

\bibitem[Hammer \etal (1997)]{ham97} Hammer, F., Flores, H., Lilly, S.J.,
Crampton, D., Rola, C., \etal, 1997, \textit{ApJ}, 481, 49  
 
\bibitem[Hogg \etal (1998)]{hog98} Hogg, D.W., Cohen, 
J.G., Blandford, R., Pahre, M.A., 1998, \textit{ApJ}, 504, 622 
 
\bibitem[Hui \etal (1995)]{hui95} Hui, X., Ford, H.C., Freeman, K.C.,
Dopita, M.A., 1995, \textit{ApJ}, 449, 592

\bibitem[Jacoby(1989)]{jac89} Jacoby, G.H. 1989, \textit{ApJ}, 339, 39

\bibitem[Jacoby, Ciardullo, Ford(1990)]{jcf90} Jacoby, G.H., Ciardullo, 
R., Ford, H.C., 1990, \textit{ApJ}, 356, 332

\bibitem[Kudritzki \etal (2000)]{kud00} Kudritzki, R.-P., Méndez, R.H.,
Feldmeier, J.J., Ciardullo, R., \etal, 2000, \textit{ApJ}, 536, 19  
 
\bibitem[M\'endez \etal (2001)]{men01} M\'endez, R.H., Riffeser, A.,
Kudritzki, R.-P., Matthias, M., \etal 2001, \textit{ApJ}, 563, 135

\bibitem[Merrett \etal (2006)]{mer06} Merrett, H.R., Merrifield, M.R.,
Douglas, N.G., \etal, \textit{MNRAS}, in press (astro-ph/0603125)

\bibitem[Mihos \etal (2005)]{mih05} Mihos, J.C., Harding, P., Feldmeier, J.,
Morrison, H., 2005, \textit{ApJ} (Letters), 631, L41

\bibitem[Peng, Ford \& Freeman(2004)]{pen04} Peng, E., Ford, H.C.,
Freeman, K.C., 2004, \textit{ApJ}, 602, 685

\bibitem[Romanowsky \etal (2003)]{rom03} Romanowsky, A.J., Douglas, N.G.,
Arnaboldi, M., Kuijken, K., \etal, 2003, \textit{Science}, 301, 1696

\bibitem[Sambhus \etal (2006)]{sam06} Sambhus, N., Gerhard, O., M\'endez, R.H., 
2006, \textit{AJ}, 131, 837

\bibitem[Teodorescu \etal (2005)]{teo05} Teodorescu, A.M., M\'endez, R.H., 
Saglia, R.P., Riffeser, A., \etal, 2005, \textit{ApJ}, 635, 290

\bibitem[Theuns \& Warren (1997)]{thw97} Theuns, T. \& Warren, 
S.J. 1997, \textit{MNRAS}, 284, L11  

\bibitem[Tran \etal (2004)]{tran+04}
Tran K.-V., Lilly, S.J., Crampton, D., Brodwin, M. 2004, 
\textit{ApJ} (Letters), 612, L89


\end{thebibliography}
\end{document}